\pdfoutput=1 
\documentclass[aps,prl,reprint,groupedaddress,amsmath,amssymb,nofootinbib
]{revtex4-1}

\usepackage{graphicx}
\usepackage[caption=false]{subfig} 
\usepackage{bm,mathtools} 
\usepackage[latin1]{inputenc} 
\usepackage{todonotes}
\usepackage{color}
\usepackage{microtype}
\newcommand{\povm}{\textsc{povm}}
\newcommand{\cpc}{\ifmmode C^+\else$C^+$\fi}
\newcommand{\cpcI}{\ifmmode\cpc_I\else$\cpc_I$\fi}

\newcommand{\PP}{\mathbb{P}}
\newcommand{\of}[1]{(#1)}
\newcommand{\vect}[1]{\mathbf{#1}}
\newcommand{\syst}[1]{\left\{ \begin{aligned} #1 \end{aligned}  \right.}    
\newcommand{\de}{\mathrm{d}}  
\usepackage{braket} 
\newcommand{\thalf}[1][1]{\tfrac{#1}{2}} 
\newcommand{\abs}[1]{\lvert#1\rvert} 
\newcommand{\Abs}[1]{\left \lvert#1\right \rvert}  
\newcommand{\intdef}[2]{\int_{\scriptscriptstyle#1}^{\scriptscriptstyle#2}\!\! } 
\newcommand{\intdefde}[3]{\intdef{#1}{#2} \de #3 \  }

\usepackage{txfonts}  

\begin{document}


\title{What does one measure, when one measures the arrival times of a quantum particle?}


\author{Nicola Vona}
\email[]{vona@math.lmu.de}
\author{G\"unter Hinrichs}
\author{Detlef D\"urr}
\affiliation{Mathematisches Institut, Ludwigs-Maximilians-Universit\"at M\"unchen, Theresienstra\ss e 39, D-80333 M\"unchen, Germany.}


\date{\today}

\begin{abstract}
What is the right statistics for the measurements of arrival times of a quantum particle?
Although this question is very old, 
it is still open.
Usual experiments are performed in far-field regime and this question becomes unimportant, as a semiclassical analysis suffices.
Nevertheless, the development in the detector technology will soon allow for near-field investigations, thus a better understanding of arrival time measurements is needed. 
Since outcomes of quantum measurements are always  described by \povm{}s,  various  arrival time \povm{}s have been proposed. On the other  hand many physicists would agree that the  arrival time statistics is given by the quantum flux.   
This  urges the question whether a \povm{} exists, which agrees approximately with the quantum flux values  on a reasonable set of wave functions. We answer this question negatively for a very natural set of wave functions, but we remark that the answer is very sensitive to the choice of the set, and provide evidence for the existence of a \povm{} that agrees with the quantum flux on a more restrictive set.
\end{abstract}


\maketitle

\section{Introduction}

\subsection{Measurement of time in quantum mechanics} 

Consider the following experiment: a one particle wave function is prepared at time zero in a certain bounded region  $G$ of space; the wave evolves freely, and  around that region are particle detectors waiting for the particle to arrive.
The times and locations at which detectors click are random, without doubts. 
We ask: What is the distribution of these random events?

The measurement of time in quantum mechanics is an old and recurrent theme, mostly because no time observable as self adjoint operator exists \cite{Pauli1958,EgusquizaMuga1999}. Time is therefore not observable in the orthodox quantum mechanical sense, but since clocks exist and  time measurements are routinely done in quantum mechanical experiments, the situation draws attention.
Usual experiments are performed in far-field regime, where a semi-classical analysis is sufficient, but with faster detectors at hand it will be soon possible to investigate the near-field regime, where a deeper analysis is needed \cite[see for example][]{ZhangSlyszVerevkin2003,PearlmanCrossSlysz2005}.

It follows easily from Born's statistical law that ordinary quantum
  measurements are described by  \povm{s},  positive operator valued measures, \cite{Ludwig1983a, DuerrGoldsteinZanghi2004,DurrGoldsteinZanghi2013}.
This fact motivated a longstanding quest for an arrival time \povm{} derived from first principles and independent of the details of the measurement interaction \cite{Kijowski1974,Werner1986,Werner1987,MugaSalaPalao1998,EgusquizaMuga1999,MugaLeavens2000}.

But what classifies an actual experiment as an arrival time measurement?
Surely not the fact that its outcomes are distributed according to a certain \povm{}, otherwise an appropriate computer program could also be called ``arrival time measurement''.
In fact,  the quest for an arrival time \povm{} cannot be grounded in the believe that there exists some  \emph{true} arrival time, the distribution of which, only because instruments readings are distributed according to a \povm{}, is conceived as a \povm{}. 
Indeed, quantum measurements in general do not  actually measure a preexisting value of an underlying quantity, and outcomes rather result from the interaction of the system with the experimental set-up.

One should rather think that any measurement that one would call arrival time measurement must necessarily satisfy some symmetry requirements, and that these requirements identify a class of \povm{}s \cite{Kijowski1974,Werner1986}.
The elements of this class correspond to different realizations of the measurement interaction, and must be treated on a case-by-case basis.

\subsection{The integral flux statistics}
In the simplified case in which the arrival position is not detected---or, similarly, if we restrict to one dimension---a general and easy analysis is possible for the initial states such that the probability that the particle is inside the region $G$ decreases monotonically with time.
To satisfy this requirement it is sufficient that the wave function of the particle belongs to the set
\begin{equation}
\cpc \coloneqq \{  \psi\:  |\ 
{\vect j}_{\psi}({\vect x},t)  \cdot \de{\vect S} \geq 0,
\ 
\forall {\vect x}\in\partial G, \ \forall t \geq  0  \}  ,
\end{equation}
where $\psi$ is the particle's wave function,  
\begin{equation}\label{fluxexpression}
{\vect j}_{\psi}({\vect x},t) 
\coloneqq
 \frac{\hbar}{m}\, \Im \left( \psi^*({\vect x},t)\, \nabla \psi({\vect x},t) \right)   
\end{equation} 
is the probability current, $\partial G$ is the boundary of $G$, and $\de{\vect S}$ is the surface element directed outwards.

In these conditions, the probability that the particle crosses  $\partial G$ later than time $t$ is equal to the probability that the particle is inside $G$ at $t$.
Therefore, the probability for an arrival at $\partial G$ during the time interval $\de t$ is given by the \emph{integral flux statistics}
\begin{equation}\label{eq:IntegralProb}
\PP (\de t) 
=
\frac{\de}{\de t} 
	\left( \int_{G} \abs{\psi(\vect x,t)}^2\,  {\de^3\vect x}  \right)
	 \, \de t
=
\left( \int_{\partial G} {\vect j}_{\psi}({\vect x},t)  \cdot \de{\vect S}\right)\, \de t  .
\end{equation}

The previous analysis, together with the fact that any quantum measurement is described by a \povm{}, raises the following question:
\begin{quote}
\emph{Does there exist a  \povm{} which agrees with the 
integral flux statistics \eqref{eq:IntegralProb} on the set \cpc{}?}
\end{quote}
We answer this question in the next sections.

\subsection{Bohmian arrival times}
The flux statistics is most naturally understood in the context of Bohmian mechanics.

In the experiment introduced above the Bohmian particle moves along the continuous trajectory $\vect X(t)$, and arrives at the detector at the time at which $\vect X(t)$ crosses it, therefore a ``true arrival time'' does exist, namely that of the Bohmian particle.

We recall that the Bohmian trajectories are the flux lines of the probability, i.e.\ 
\begin{equation}\label{BM}
\dot{\vect X}(t)
=\frac{{\vect j}_{\psi}\bm({\vect X}(t),t\bm)}{|\psi\bm({\vect X}(t),t)\bm)|^2}  .
\end{equation}
The particle's wave function in Eq.\ \eqref{BM} can in principle also be the so called conditional wave function, which takes into account the interaction with the detector \cite{DuerrGoldsteinZanghi1992,DurrGoldsteinZanghi2013,PladevallOriolsMompart2012}.
A Bohmian particle can in general cross the surface $\partial G$ several times and the probability for having a first arrival at the surface element $\de \vect S$ during the time interval $\de t$ is
\begin{equation}
\label{POPtruncatedflux}
\PP (\de \vect S, \de t) = 
\tilde{\vect j}_{\psi}  \cdot \de{\vect S}\: \de t  ,
\end{equation}
where 
\begin{equation}
\label{truncatedflux}
 { \tilde{\vect j}}_{\psi}({\vect x},t)
 \coloneqq
 \syst{ & {\vect j}_{\psi}({\vect x},t) && \mbox{if $(t,{\vect x})$ is a first exit from
$G$}  
\\ &0&& \mbox{otherwise} }  
\end{equation}
is the so called truncated current \cite{DaumerDurrGoldstein1997,DurrGoldsteinZanghi2013}.
A first exit event $(t,\vect x)$  from the region $G$ is such that the Bohmian trajectory crosses $\partial G$ for the first time since $t=0$ through the point $\vect x \in \partial G$ at time $t$.

In case each Bohmian trajectory crosses the detector surface only once---i.e.\ the wave function belongs to the set $\cpc$---then every exit is a first exit, and  the first arrival statistics is given by the simpler expression
\begin{equation}
\label{POPRAN}
\PP (\de \vect S, \de t) = 
{\vect j}_{\psi}  \cdot \de{\vect S}\: \de t,
\end{equation}
which we shall call the \emph{flux statistics}.
Note that this gives the statistics for both arrival time and arrival position.

Now one may ask if it is possible to design an experiment whose results disclose the  ``true arrival times''.
The outcomes of such an experiment would be distributed according to Eq.~\eqref{POPtruncatedflux}.
 Unfortunately, this is  impossible, since the truncated flux depends explicitly on the trajectory of the particle, and is not sesquilinear with respect to the wave function as needed for a \povm{} \cite[see also][]{RuggenthalerGrublKreidl2005}.
Hence, according to Bohmian mechanics the ``true arrival time'' exists, but its statistic is not given by a \povm{}, so there is no experiment able to measure it (note that this statement is not in contradiction with the fact that the Bohmian trajectories and the quantum flux are detectable  in weak measurements \cite{Wiseman2007,KocsisBravermanRavets2011,TraversaAlbaredaDi-Ventra2012}).
From this circumstance one may  jump to the conclusion that Bohmian mechanics must be false. 
That conclusion is however unwarranted. 
The measurement analysis in Bohmian mechanics yields straightforwardly that the statistics of measurement outcomes are always given by \povm{}s \cite{DuerrGoldsteinZanghi2004,DurrGoldsteinZanghi2013}. There is no inconsistency here.
Observing that a \povm{} is defined on the whole of the Hilbert space, we see that our previous request of measurability was rather strong, in that we allowed any initial state for the particle, even the very bizarre ones.
As a consequence, it is reasonable to restrict our quest for measurability to a subset of  good wave functions, as for example \cpc{}.
Now we may ask the following question:
\begin{quote}
\emph{Does there exist a  \povm{} which agrees with the 
flux statistics \eqref{POPRAN} on the set \cpc{}?}
\end{quote}
This question slightly generalizes that asked in the previous section.

\section{No go theorem for the arrival time POVM}
For simplicity we consider a particle moving in one dimension with a detector only at one place.
That restricts our analysis to random times only, and makes \eqref{eq:IntegralProb} and \eqref{POPRAN} equivalent, which is sufficient for the purpose at hand; the generalization to three dimensions is straightforward.
We consider that the detector is placed at $D>0$ and that it is active  during the time interval $I=(0, T)$.
The one particle wave is prepared at time zero well located around the origin.

We introduce the set of wave functions
\begin{equation}
\cpcI \coloneqq \{  \psi\:  |\ 
{ j}_{\psi}(D,t) \geq  0,
\ 
\forall t \in I  \}  .
\end{equation}
On these wave functions the flux statistics is the first arrival time statistics.
We want to find out if a \povm{} density $ O_t$ exists, such that
\begin{equation}\label{eq:JPOVM}
\braket{ \psi |  O_t  |  \psi }
= j_{\psi}(D,t)
\quad \forall t \in I,\ \  
 \forall \psi \in \cpcI  .
\end{equation}
In the following we will use the notation
\begin{equation}
j_+ \coloneqq  j_{\psi+\varphi} ,
\qquad
j_- \coloneqq  j_{\psi-\varphi} .
\end{equation}
By sesquilinearity   of \eqref{fluxexpression} we have
\begin{equation}
\label{eq:JLinearity}
j_\psi + j_\varphi  =  \thalf  \bigl(j_+  +  j_- \bigr)  .
\end{equation}
Similarly,
\begin{multline}
\label{eq:POVMLinearity}
\braket{ \psi |  O_t  |  \psi }  +  \braket{ \varphi |  O_t  |  \varphi } = \\
	=  \thalf \bigl(  \braket{ \psi + \varphi |  O_t  |  \psi + \varphi }   
		+  \braket{ \psi - \varphi |  O_t  |  \psi - \varphi } \bigr)  .
\end{multline}

\begin{figure*}
\subfloat[\label{subfig:InterferenceSumDifferencePlus}]{\includegraphics[width=.71\columnwidth]{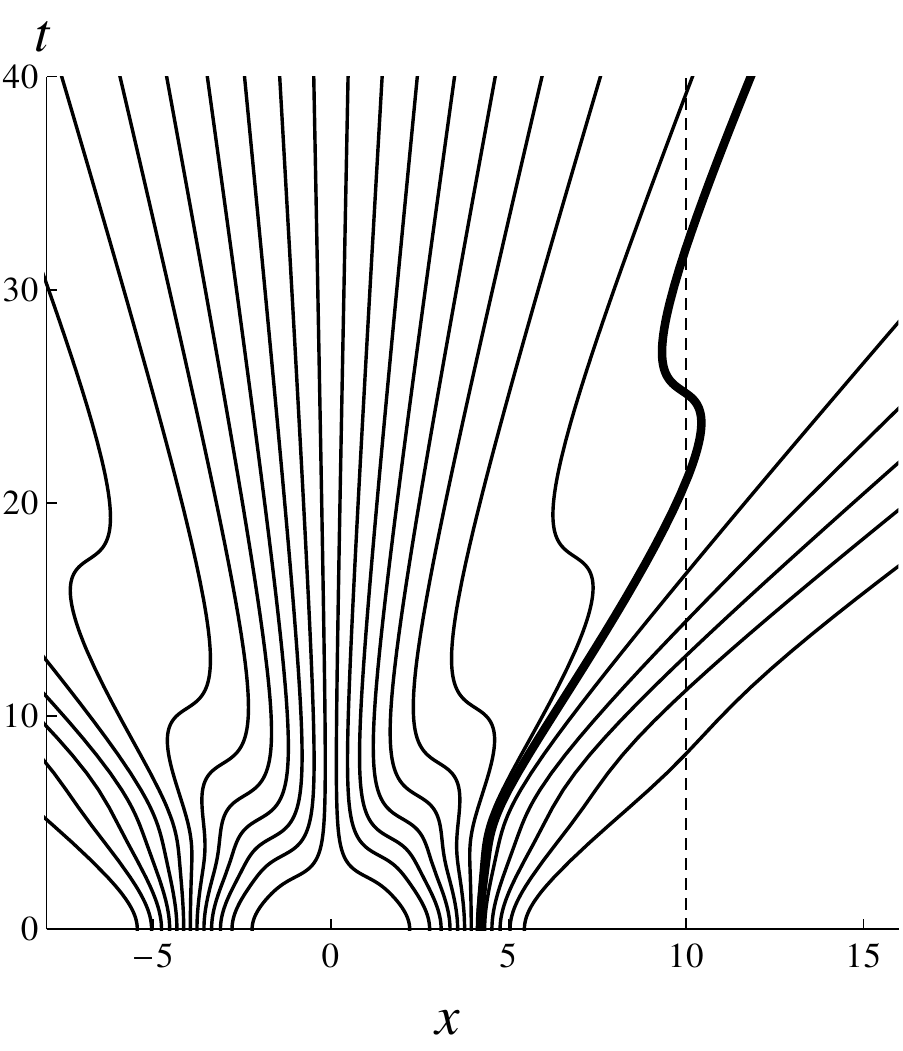}}  
\hfil
\subfloat[\label{subfig:InterferenceSumDifferenceMinus}]{\includegraphics[width=.71
\columnwidth]{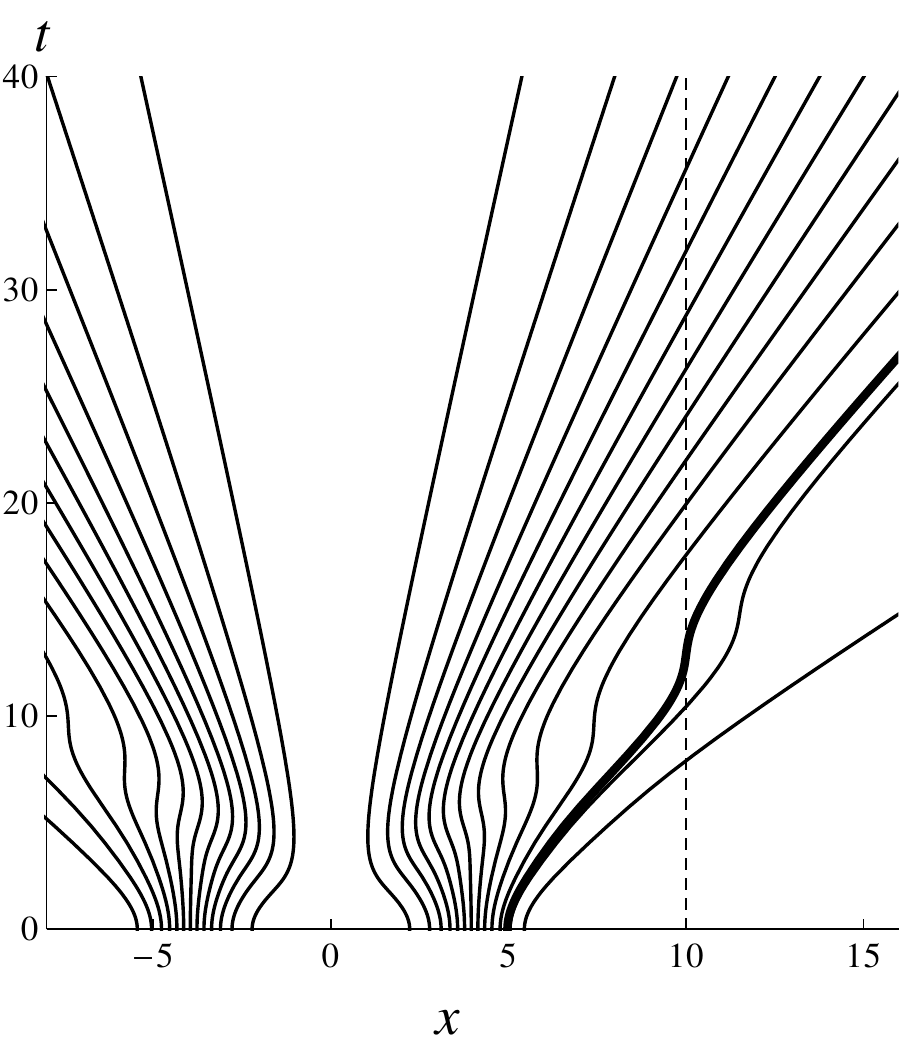}}  
\caption[]{ 
Bohmian trajectories for \subref{subfig:InterferenceSumDifferencePlus} $g_1+g_2$ ($\notin\cpcI$),  and \subref{subfig:InterferenceSumDifferenceMinus} $g_1-g_2$ ($\in\cpcI$), with $g_{1,2}$ Gaussian packets with unitary  position variance, zero mean momentum, and initial mean position equal to $4$ and $-4$, respectively. 
The detector is located at $D=10$ (dashed line).
The initial positions are distributed according to $|\psi|^2$; one further trajectory is shown (bold line), such that it crosses the screen at a minimum of the current.
The units are such that $m=\hbar=1$.
\label{fig:InterferenceSumDifference}}
\end{figure*}

Consider now two wave functions $\psi$ and $\varphi$ in $\cpcI$  such that also $\psi+\phi$ is in $\cpcI$, while $j_-(D, t_-)<0 $ for some $t_-\in I$. 
Such functions exist, and an example built with Gaussian wave packets is given in Fig.\ \ref{fig:InterferenceSumDifference}.
Requiring \eqref{eq:JPOVM}, we have for every $t$ in $I$ (omitting the argument $D$ in $j$)
\begin{gather}
\braket{ \psi |  O_t  |  \psi } =  j_\psi\of{t},
\qquad
\braket{ \varphi |  O_t  |  \varphi } =  j_\varphi\of{t},
\nonumber\\
\text{and }
\braket{ \psi + \varphi |  O_t  |  \psi + \varphi } =  j_+\of{t}  .
\end{gather}
Substituting in \eqref{eq:JLinearity} and using \eqref{eq:POVMLinearity} we thus get
\begin{equation}
\braket{ \psi - \varphi |  O_t  |  \psi - \varphi }
= j_-\of{t}
\quad \forall t \in I\,.
\end{equation}
But $\braket{ \psi - \varphi |  O_t  |  \psi - \varphi } $ is positive for all $t$ in $I$, while $j_- $ becomes negative at $t_- \in I$, hence a contradiction.
Therefore, \emph{a \povm{} satisfying \eqref{eq:JPOVM} on all functions in \cpcI{} does not exist}.

We can strengthen the result.
Let $O_\psi(t) \coloneqq \braket{ \psi |  O_t  |  \psi }$, 
$\epsilon_\psi\of{t} \coloneqq O_\psi(t) - j_\psi(t)$, 
and let $\epsilon_\pm \coloneqq O_{\psi\pm\varphi}(t) - j_\pm(t)$.
By linearity, i.e.\ subtracting Eq.~\eqref{eq:JLinearity} and \eqref{eq:POVMLinearity},
\begin{equation}
\epsilon_\psi + \epsilon_\varphi  =  \thalf  (\epsilon_+  +  \epsilon_-)  ,
\end{equation}
that implies
\begin{equation}
2\abs{\epsilon_\psi} + 2\abs{\epsilon_\varphi} + \abs{\epsilon_+}
	\geq\abs{2\epsilon_\psi + 2\epsilon_\varphi  - \epsilon_+}
	= \abs{ \epsilon_-}  .
\end{equation}
At a time $t_-$ such that $j_-\of{t_-}<0$, we have $\abs{\epsilon_-\of{t_-}}>\abs{j_-\of{t_-}}$ and thus
\begin{equation}
2\abs{\epsilon_\psi\of{t_-}} + 2\abs{\epsilon_\varphi\of{t_-}} + \abs{\epsilon_+\of{t_-}}
	>\abs{j_-\of{t_-}}  .
\end{equation}
The value $\abs{j_-\of{t_-}}$ is in general not bounded, therefore the error
between any \povm{} and the flux statistics can be arbitrarily large.
 The conclusion is therefore that \emph{there exists no \povm{} which approximates the flux statistics on all functions in \cpcI{}}.

\subsection{The argument is a set argument}
We wish to stress that in the previous section we showed that it is impossible to design an experiment that measures the Bohmian arrival time \emph{on all wave functions in a certain set}, namely \cpcI.
The choice of the set that we consider is crucial, and on a different set our argument may not apply.
To illustrate this point, we present an exaggerated example.
Consider the set of wave functions
\begin{equation}
\cpc_\mathrm G
\coloneqq \{  \psi \in\cpcI \:  |\ 
\psi \text{ is a Gaussian}  \}  .
\end{equation}
For every $\psi$ and $\phi$ in $\cpc_\mathrm G$, neither $\psi + \phi$ nor $\psi - \phi$ is in $\cpc_\mathrm G$, and our argument does not apply.
Of course, the set $\cpc_\mathrm G$ is absolutely artificial and serves only to highlight that our impossibility result depends heavily on the choice of the class of allowed wave functions.

\section{Scattering states}
A class of functions very important from the experimental point of view is that of scattering states, i.e.\ states that reach the detector in far field regime.
These wave functions are particularly important because usual time measurements  are performed in these conditions.
For these states \cite{BrenigHaag1959,Dollard1969} (in units such that $m=\hbar=1$)
\begin{equation}\label{eq:ScatteringState}
\psi_{\!t}(x)
\approx \frac  {e^{i x^2 /2t}}   {(it)^{1/2}}\ \,
	\tilde\psi \bigl(\tfrac{x}{t} \bigr),
\qquad 
x\approx D,\ 
\forall t \in I  ,
\end{equation}
where $\tilde\psi$ is the Fourier transform of the initial wave function.
As a consequence, it can be shown that \cite{DurrTeufel2009}
\begin{equation}\label{eq:ScatteringCurrent}
j_{\psi}(D,t)
\approx \frac{D}{t^2}\,
	\Abs{\tilde\psi \bigl(\tfrac{D}{t} \bigr)}^2 ,
\qquad \forall t \in I  ,
\end{equation}
and therefore all scattering states are in \cpcI.
A linear combination of scattering states is still a scattering state, indeed Eq.~\eqref{eq:ScatteringState} and \eqref{eq:ScatteringCurrent} apply to the combination as well.
Therefore, no contradiction arises asking for a \povm{} that agrees with the flux statistics on scattering states.
An example of such a \povm, at least approximately, is given by the momentum operator.
This follows from Eq.~\eqref{eq:ScatteringCurrent}, that shows that Bohmian arrival time measurements on scattering states are nothing else than momentum measurements.
In conclusion, our negative result about \cpcI{}  does not forbid to interpret actual, far field time measurements in terms of the flux statistics.

\section{High-energy wave functions}
As already remarked, the set for which we ask accordance between the flux statistics and the \povm{} is crucial.
We found that the set of scattering states presents no problem, but it would be of course much more interesting to identify a subset of \cpcI{}, such that it is possible to measure the Bohmian arrival time also in near field conditions.
We do not have any proof that such a set exists, nevertheless we believe that the subset of \cpcI{} of wave functions with high energy is a good candidate, at least in an approximate sense.

To support our conjecture, we performed some numerical investigations.%
\footnote{See the Appendix for more details. 
Our conjecture is supported also by the results of \cite{YearsleyDownsHalliwell2011} for a wide class of clock models.}
We considered as model system the superposition of two Gaussian packets $g_1$ and $g_2$, with equal position variance $\sigma$.
If $g_1$ and $g_2$ are both elements of \cpcI, then the eventual negative current of their superpositions must be caused by interference, that is in turn either due to the spreading of the packets, or to their different velocities.

To study the effect of the spreading, we first set the mean momenta of the two packets to be the same, and equal to $k$.
Varying $\sigma$, we found that a threshold $k_\sigma$ exists, such that for $k$ smaller than $k_\sigma$ it is possible that $g_1+g_2$ is in $\cpcI$, but $g_1-g_2$ is not, while for $k$ larger than $k_\sigma$ both $g_1+g_2$ and $g_1-g_2$ are in $\cpcI$.
The threshold $k_\sigma$ increases with decreasing $\sigma$, as expected from the fact that a smaller $\sigma$ means a larger momentum variance, and therefore a larger probability of small momentum.

We examined the effect of a difference in the velocities of the two packets considering the closest packet to the screen to have a fixed momentum $k_1$ well above the value $k_\sigma$, and varying the momentum $k_2$ of the second packet. 
We found that, if $k_2$ is sufficiently far away from $k_1$, then neither $g_1+g_2$ nor $g_1-g_2$ is in \cpcI.
On the contrary, for $k_2$ close to $k_1$, it can happen that the sum is in \cpcI{} and the difference is not, or the other way round.
However, the interval {of $k_2$ values around $k_1$ for } which this happens shrinks (relatively to $k_1$) with growing $k_1$, as well as the maximal value of the negative current.

For the subset of \cpcI{} of  wave functions with high energy it is therefore not true that it is possible to find a \povm{} that agrees \emph{exactly} with the flux statistics, indeed our main argument still applies.
Nevertheless, our numerical study supports the conjecture that it is possible to find  a \povm{} that \emph{approximately} agrees with the flux statistics, with a better agreement for higher energies.

\section{Conclusions}
We showed that  \emph{no \povm{} exists, that approximates the flux statistics on all  functions in \cpcI{}}.
Moreover, the error $\epsilon_\psi$ between a candidate \povm{} and the flux statistic can be very large on any  wave function in \cpcI{}, even for simple states like Gaussians or sum of Gaussians.

This negative result is very sensitive to the choice of the set; for example it is  possible to find a \povm{} that agrees with the flux statistics on the subset of \cpcI{} composed by scattering states.
Similarly, we conjecture that a \povm{} exists, that approximates the flux statistics on the subset of \cpcI{} of wave functions with high energy. 
We produced some numerical evidence to support this conjecture.

\begin{acknowledgments}
The authors are grateful to Lee Rozema for suggesting some references.
This work benefitted from the COST action ``Fundamental Problems in Quantum Physics".
Nicola Vona gratefully acknowledges the financial support of the Elite Network of Bavaria.
\end{acknowledgments}

\section{Appendix}

We present here the numerical calculations that we performed to investigate if a \povm{} can exist, that agrees with the flux statistics  on the subset of \cpcI{} of wave functions with high energy.
Our model system was the superposition of two Gaussian packets $g_1$ and $g_2$ with equal initial position variance $\sigma$.
We used units are such that $m=\hbar=1$, and we considered the time interval $I=(0,T)$, where 
\begin{equation}
T=\frac {x_2-3\sigma}    {k_2+5/12}  ,
\end{equation}
$x_2$ is the initial mean position of the furthest packet from the screen, and $k_2$ is its mean momentum; the term $5/12$ has been inserted by hand to ensure that $T$ is reasonably small also when $k_2$ is zero.

\subsection{Effect of the spreading}

\begin{figure}
\subfloat[\label{subfig:Sigma1}]{\includegraphics[width=.45\columnwidth]{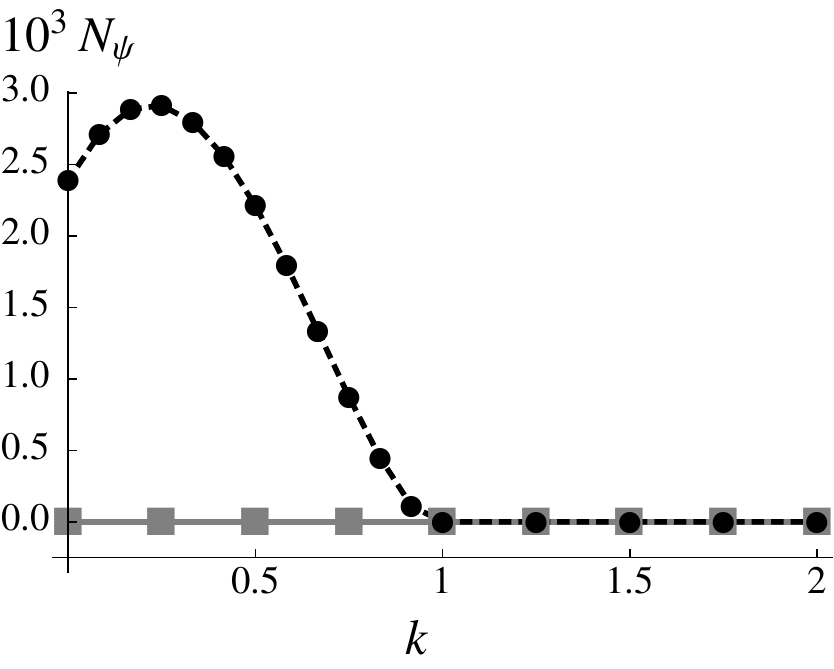}}  
\hfil
\subfloat[\label{subfig:Threshold}]{\includegraphics[width=.45\columnwidth]{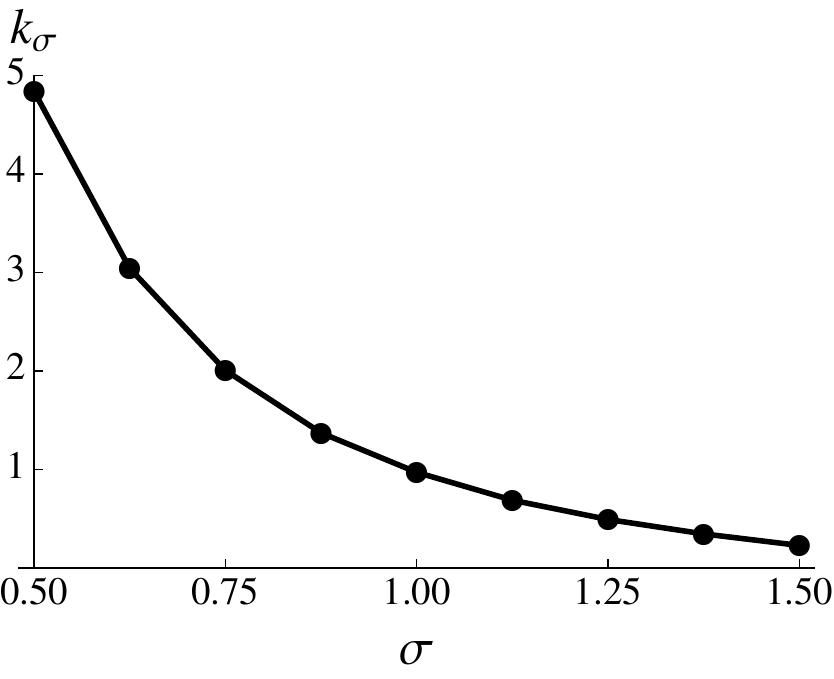}}  
\caption[]{ 
\subref{subfig:Sigma1} 
Numerically integrated negative current $N_\psi$ for $\psi=g_1+g_2$ (circles, dashed black line) and for $\psi=g_1-g_2$ (squares, solid gray line), as functions of the mean momentum $k$ of the two packets.
\subref{subfig:Threshold} 
Threshold $k_\sigma$ as a function of the position variance $\sigma$ of the two packets.
\label{fig:SpreadingIntegral}}
\end{figure}

We studied the effect of the spreading setting the mean momenta of the two packets to be both equal to $k$.
We quantified the total amount of negative current during the interval $I$ by
\begin{align}
N_\psi &\coloneqq
\intdefde{I}{}{t}
| j_{\psi}(D,t)| \
\chi_< (t),
\nonumber\\
&\text{with}\quad\chi_< (t) \coloneqq
\syst{1, &&&j_{\psi}(D,t)<0
\\
0, &&&j_{\psi}(D,t)\geq0  .}  
\end{align}
In Fig.\ \ref{subfig:Sigma1} we plotted $N_{g_1+g_2}$ and $N_{g_1-g_2}$ as functions of $k$, for two Gaussian packets with unitary  position variance, zero mean momentum, and initial mean position equal to $4$ and $-4$, respectively; the detector is located at $D=10$. 
For $k$ bigger than one no negative current is present, and both $g_1+g_2$ and $g_1-g_2$ are in \cpcI{}.

We denoted this threshold value by $k_\sigma$, and we found that it decreases as $\sigma$ increases, as shown in Fig.\ \ref{subfig:Threshold}.

\subsection{Effect of different velocities}

\begin{figure}
\includegraphics[width=.72\columnwidth]{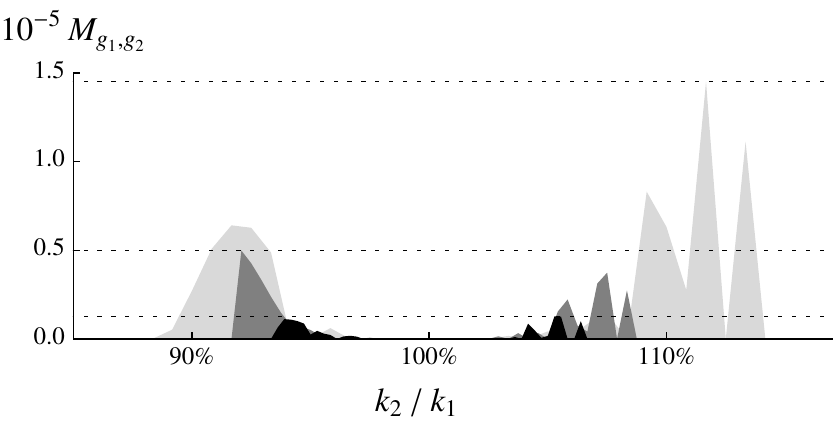}  
\caption{
Numerical values of $M_{g_1,g_2}$  as a function of the ratio of the momenta of the two packets, for  $k_1 = 20$  (light gray), $40$ (gray), and $60$ (black). 
The horizontal dashed lines highlight the maximal values of $M_{g_1,g_2}$.
\label{fig:DifferentVelocitiesIntegral}}
\end{figure}

To study the effect of a difference in the velocities of the two packets we considered $g_1$ and $g_2$ to have again unitary position variance, but different mean momenta $k_1$ and $k_2$, respectively. 
The packet $g_1$ was initially centered around zero, and we considered the values $20$, $40$, and $60$ for its mean momentum $k_1$.
The initial mean position $x_2$ of the packet $g_2$ was such that the two maxima cross in $D+\sigma_t$, where $\sigma_t$ is the position variance at the time of the crossing, and $D=40$ is the detector position.
Consequently, $x_2$ ranged approximately between $-20$ and $-80$, depending on $k_2$.

We studied the quantity
\begin{equation}
M_{g_1,g_2} \coloneqq
(  N_{g_1+g_2} + N_{g_1-g_2}  )\  \chi_0  ,
\end{equation}
where
\begin{equation}
\chi_0 \coloneqq
\syst{0, &&&\text{if either $N_{g_1+g_2}$ or $N_{g_1-g_2}$ is zero,}
\\&&&\text{but not both of them,}
\\
1, &&&\text{otherwise} .}  
\end{equation}
Therefore,  $M_{g_1,g_2}$ is zero when both $g_1+g_2$ and $g_1-g_2$ are in \cpcI{}, as well as when none of them is, while $M_{g_1,g_2}$ is different from zero when one combination is in \cpcI{} and the other one is not.
The results are presented in Fig.\ \ref{fig:DifferentVelocitiesIntegral}, from which it is evident that the interval on which $M_{g_1,g_2}$ is different from zero narrows (relatively to $k_1$) with growing $k_1$, and at the same time the maximal value of $M_{g_1,g_2}$ decreases.

\bibliography{biblio.bib}

\begin{thebibliography}{23}%
\makeatletter
\providecommand \@ifxundefined [1]{%
 \@ifx{#1\undefined}
}%
\providecommand \@ifnum [1]{%
 \ifnum #1\expandafter \@firstoftwo
 \else \expandafter \@secondoftwo
 \fi
}%
\providecommand \@ifx [1]{%
 \ifx #1\expandafter \@firstoftwo
 \else \expandafter \@secondoftwo
 \fi
}%
\providecommand \natexlab [1]{#1}%
\providecommand \enquote  [1]{``#1''}%
\providecommand \bibnamefont  [1]{#1}%
\providecommand \bibfnamefont [1]{#1}%
\providecommand \citenamefont [1]{#1}%
\providecommand \href@noop [0]{\@secondoftwo}%
\providecommand \href [0]{\begingroup \@sanitize@url \@href}%
\providecommand \@href[1]{\@@startlink{#1}\@@href}%
\providecommand \@@href[1]{\endgroup#1\@@endlink}%
\providecommand \@sanitize@url [0]{\catcode `\\12\catcode `\$12\catcode
  `\&12\catcode `\#12\catcode `\^12\catcode `\_12\catcode `\%12\relax}%
\providecommand \@@startlink[1]{}%
\providecommand \@@endlink[0]{}%
\providecommand \url  [0]{\begingroup\@sanitize@url \@url }%
\providecommand \@url [1]{\endgroup\@href {#1}{\urlprefix }}%
\providecommand \urlprefix  [0]{URL }%
\providecommand \Eprint [0]{\href }%
\providecommand \doibase [0]{http://dx.doi.org/}%
\providecommand \selectlanguage [0]{\@gobble}%
\providecommand \bibinfo  [0]{\@secondoftwo}%
\providecommand \bibfield  [0]{\@secondoftwo}%
\providecommand \translation [1]{[#1]}%
\providecommand \BibitemOpen [0]{}%
\providecommand \bibitemStop [0]{}%
\providecommand \bibitemNoStop [0]{.\EOS\space}%
\providecommand \EOS [0]{\spacefactor3000\relax}%
\providecommand \BibitemShut  [1]{\csname bibitem#1\endcsname}%
\let\auto@bib@innerbib\@empty
\bibitem [{\citenamefont {Pauli}(1958)}]{Pauli1958}%
  \BibitemOpen
  \bibfield  {author} {\bibinfo {author} {\bibfnamefont {W.}~\bibnamefont
  {Pauli}},\ }in\ \href@noop {} {\emph {\bibinfo {booktitle} {Encyclopedia of
  Physics}}},\ Vol.\ \bibinfo {volume} {5/1},\ \bibinfo {editor} {edited by\
  \bibinfo {editor} {\bibfnamefont {S.}~\bibnamefont {Flugge}}}\ (\bibinfo
  {publisher} {Springer, Berlin},\ \bibinfo {year} {1958})\ p.~\bibinfo {pages}
  {60}\BibitemShut {NoStop}%
\bibitem [{\citenamefont {Egusquiza}\ and\ \citenamefont
  {Muga}(1999)}]{EgusquizaMuga1999}%
  \BibitemOpen
  \bibfield  {author} {\bibinfo {author} {\bibfnamefont {I.~L.}\ \bibnamefont
  {Egusquiza}}\ and\ \bibinfo {author} {\bibfnamefont {J.~G.}\ \bibnamefont
  {Muga}},\ }\href {\doibase 10.1103/PhysRevA.61.012104} {\bibfield  {journal}
  {\bibinfo  {journal} {Phys. Rev. A}\ }\textbf {\bibinfo {volume} {61}},\
  \bibinfo {pages} {012104} (\bibinfo {year} {1999})}\BibitemShut {NoStop}%
\bibitem [{\citenamefont {Zhang}\ \emph {et~al.}(2003)\citenamefont {Zhang},
  \citenamefont {Slysz}, \citenamefont {Verevkin}, \citenamefont {Okunev},
  \citenamefont {Chulkova}, \citenamefont {Korneev}, \citenamefont {Lipatov},
  \citenamefont {Gol'tsman},\ and\ \citenamefont
  {Sobolewski}}]{ZhangSlyszVerevkin2003}%
  \BibitemOpen
  \bibfield  {author} {\bibinfo {author} {\bibfnamefont {J.}~\bibnamefont
  {Zhang}}, \bibinfo {author} {\bibfnamefont {W.}~\bibnamefont {Slysz}},
  \bibinfo {author} {\bibfnamefont {A.}~\bibnamefont {Verevkin}}, \bibinfo
  {author} {\bibfnamefont {O.}~\bibnamefont {Okunev}}, \bibinfo {author}
  {\bibfnamefont {G.}~\bibnamefont {Chulkova}}, \bibinfo {author}
  {\bibfnamefont {A.}~\bibnamefont {Korneev}}, \bibinfo {author} {\bibfnamefont
  {A.}~\bibnamefont {Lipatov}}, \bibinfo {author} {\bibfnamefont
  {G.}~\bibnamefont {Gol'tsman}}, \ and\ \bibinfo {author} {\bibfnamefont
  {R.}~\bibnamefont {Sobolewski}},\ }\href {\doibase 10.1109/TASC.2003.813675}
  {\bibfield  {journal} {\bibinfo  {journal} {Applied Superconductivity, IEEE
  Transactions on}\ }\textbf {\bibinfo {volume} {13}},\ \bibinfo {pages} {180}
  (\bibinfo {year} {2003})}\BibitemShut {NoStop}%
\bibitem [{\citenamefont {Pearlman}\ \emph {et~al.}(2005)\citenamefont
  {Pearlman}, \citenamefont {Cross}, \citenamefont {Slysz}, \citenamefont
  {Zhang}, \citenamefont {Verevkin}, \citenamefont {Currie}, \citenamefont
  {Korneev}, \citenamefont {Kouminov}, \citenamefont {Smirnov}, \citenamefont
  {Voronov}, \citenamefont {Gol'tsman},\ and\ \citenamefont
  {Sobolewski}}]{PearlmanCrossSlysz2005}%
  \BibitemOpen
  \bibfield  {author} {\bibinfo {author} {\bibfnamefont {A.}~\bibnamefont
  {Pearlman}}, \bibinfo {author} {\bibfnamefont {A.}~\bibnamefont {Cross}},
  \bibinfo {author} {\bibfnamefont {W.}~\bibnamefont {Slysz}}, \bibinfo
  {author} {\bibfnamefont {J.}~\bibnamefont {Zhang}}, \bibinfo {author}
  {\bibfnamefont {A.}~\bibnamefont {Verevkin}}, \bibinfo {author}
  {\bibfnamefont {M.}~\bibnamefont {Currie}}, \bibinfo {author} {\bibfnamefont
  {A.}~\bibnamefont {Korneev}}, \bibinfo {author} {\bibfnamefont
  {P.}~\bibnamefont {Kouminov}}, \bibinfo {author} {\bibfnamefont
  {K.}~\bibnamefont {Smirnov}}, \bibinfo {author} {\bibfnamefont
  {B.}~\bibnamefont {Voronov}}, \bibinfo {author} {\bibfnamefont
  {G.}~\bibnamefont {Gol'tsman}}, \ and\ \bibinfo {author} {\bibfnamefont
  {R.}~\bibnamefont {Sobolewski}},\ }\href {\doibase 10.1109/TASC.2005.849926}
  {\bibfield  {journal} {\bibinfo  {journal} {Applied Superconductivity, IEEE
  Transactions on}\ }\textbf {\bibinfo {volume} {15}},\ \bibinfo {pages} {579}
  (\bibinfo {year} {2005})}\BibitemShut {NoStop}%
\bibitem [{\citenamefont {Ludwig}(1985)}]{Ludwig1983a}%
  \BibitemOpen
  \bibfield  {author} {\bibinfo {author} {\bibfnamefont {G.}~\bibnamefont
  {Ludwig}},\ }\href@noop {} {\emph {\bibinfo {title} {Foundations of quantum
  mechanics}}},\ Texts and monographs in physics\ (\bibinfo  {publisher}
  {Springer-Verlag},\ \bibinfo {year} {1983,1985})\BibitemShut {NoStop}%
\bibitem [{\citenamefont {{D{{\"u}}rr}}\ \emph {et~al.}(2004)\citenamefont
  {{D{{\"u}}rr}}, \citenamefont {{Goldstein}},\ and\ \citenamefont
  {{Zangh{\`\i}}}}]{DuerrGoldsteinZanghi2004}%
  \BibitemOpen
  \bibfield  {author} {\bibinfo {author} {\bibfnamefont {D.}~\bibnamefont
  {{D{{\"u}}rr}}}, \bibinfo {author} {\bibfnamefont {S.}~\bibnamefont
  {{Goldstein}}}, \ and\ \bibinfo {author} {\bibfnamefont {N.}~\bibnamefont
  {{Zangh{\`\i}}}},\ }\href {\doibase 10.1023/B:JOSS.0000037234.80916.d0}
  {\bibfield  {journal} {\bibinfo  {journal} {Journal of Statistical Physics}\
  }\textbf {\bibinfo {volume} {116}},\ \bibinfo {pages} {959} (\bibinfo {year}
  {2004})},\ \Eprint {http://arxiv.org/abs/arXiv:quant-ph/0308038}
  {arXiv:quant-ph/0308038} \BibitemShut {NoStop}%
\bibitem [{\citenamefont {D{\"u}rr}\ \emph {et~al.}(2013)\citenamefont
  {D{\"u}rr}, \citenamefont {Goldstein},\ and\ \citenamefont
  {Zangh{\`\i}}}]{DurrGoldsteinZanghi2013}%
  \BibitemOpen
  \bibfield  {author} {\bibinfo {author} {\bibfnamefont {D.}~\bibnamefont
  {D{\"u}rr}}, \bibinfo {author} {\bibfnamefont {S.}~\bibnamefont {Goldstein}},
  \ and\ \bibinfo {author} {\bibfnamefont {N.}~\bibnamefont {Zangh{\`\i}}},\
  }\href {http://books.google.de/books?id=LpnwCJfU-54C} {\emph {\bibinfo
  {title} {Quantum Physics Without Quantum Philosophy}}}\ (\bibinfo
  {publisher} {Springer},\ \bibinfo {year} {2013})\BibitemShut {NoStop}%
\bibitem [{\citenamefont {Kijowski}(1974)}]{Kijowski1974}%
  \BibitemOpen
  \bibfield  {author} {\bibinfo {author} {\bibfnamefont {J.}~\bibnamefont
  {Kijowski}},\ }\href {\doibase 10.1016/S0034-4877(74)80004-2} {\bibfield
  {journal} {\bibinfo  {journal} {Reports on Mathematical Physics}\ }\textbf
  {\bibinfo {volume} {6}},\ \bibinfo {pages} {361 } (\bibinfo {year}
  {1974})}\BibitemShut {NoStop}%
\bibitem [{\citenamefont {Werner}(1986)}]{Werner1986}%
  \BibitemOpen
  \bibfield  {author} {\bibinfo {author} {\bibfnamefont {R.}~\bibnamefont
  {Werner}},\ }\href {\doibase 10.1063/1.527184} {\bibfield  {journal}
  {\bibinfo  {journal} {Journal of Mathematical Physics}\ }\textbf {\bibinfo
  {volume} {27}},\ \bibinfo {pages} {793} (\bibinfo {year} {1986})}\BibitemShut
  {NoStop}%
\bibitem [{\citenamefont {Werner}(1987)}]{Werner1987}%
  \BibitemOpen
  \bibfield  {author} {\bibinfo {author} {\bibfnamefont {R.}~\bibnamefont
  {Werner}},\ }\href@noop {} {\bibfield  {journal} {\bibinfo  {journal}
  {Annales de l'institut Henri Poincar{\'e} (A) Physique th{\'e}orique}\
  }\textbf {\bibinfo {volume} {47}},\ \bibinfo {pages} {429} (\bibinfo {year}
  {1987})}\BibitemShut {NoStop}%
\bibitem [{\citenamefont {Muga}\ \emph {et~al.}(1998)\citenamefont {Muga},
  \citenamefont {Sala},\ and\ \citenamefont {Palao}}]{MugaSalaPalao1998}%
  \BibitemOpen
  \bibfield  {author} {\bibinfo {author} {\bibfnamefont {J.}~\bibnamefont
  {Muga}}, \bibinfo {author} {\bibfnamefont {R.}~\bibnamefont {Sala}}, \ and\
  \bibinfo {author} {\bibfnamefont {J.}~\bibnamefont {Palao}},\ }\href
  {\doibase 10.1006/spmi.1997.0544} {\bibfield  {journal} {\bibinfo  {journal}
  {Superlattices and Microstructures}\ }\textbf {\bibinfo {volume} {23}},\
  \bibinfo {pages} {833 } (\bibinfo {year} {1998})}\BibitemShut {NoStop}%
\bibitem [{\citenamefont {Muga}\ and\ \citenamefont
  {Leavens}(2000)}]{MugaLeavens2000}%
  \BibitemOpen
  \bibfield  {author} {\bibinfo {author} {\bibfnamefont {J.}~\bibnamefont
  {Muga}}\ and\ \bibinfo {author} {\bibfnamefont {C.}~\bibnamefont {Leavens}},\
  }\href {\doibase 10.1016/S0370-1573(00)00047-8} {\bibfield  {journal}
  {\bibinfo  {journal} {Physics Reports}\ }\textbf {\bibinfo {volume} {338}},\
  \bibinfo {pages} {353 } (\bibinfo {year} {2000})}\BibitemShut {NoStop}%
\bibitem [{\citenamefont {{D{{\"u}}rr}}\ \emph {et~al.}(1992)\citenamefont
  {{D{{\"u}}rr}}, \citenamefont {{Goldstein}},\ and\ \citenamefont
  {{Zangh{\`\i}}}}]{DuerrGoldsteinZanghi1992}%
  \BibitemOpen
  \bibfield  {author} {\bibinfo {author} {\bibfnamefont {D.}~\bibnamefont
  {{D{{\"u}}rr}}}, \bibinfo {author} {\bibfnamefont {S.}~\bibnamefont
  {{Goldstein}}}, \ and\ \bibinfo {author} {\bibfnamefont {N.}~\bibnamefont
  {{Zangh{\`\i}}}},\ }\href {\doibase 10.1007/BF01049004} {\bibfield  {journal}
  {\bibinfo  {journal} {Journal of Statistical Physics}\ }\textbf {\bibinfo
  {volume} {67}},\ \bibinfo {pages} {843} (\bibinfo {year} {1992})},\ \Eprint
  {http://arxiv.org/abs/arXiv:quant-ph/0308039} {arXiv:quant-ph/0308039}
  \BibitemShut {NoStop}%
\bibitem [{\citenamefont {Pladevall}\ \emph {et~al.}(2012)\citenamefont
  {Pladevall}, \citenamefont {Oriols},\ and\ \citenamefont
  {Mompart}}]{PladevallOriolsMompart2012}%
  \BibitemOpen
  \bibfield  {author} {\bibinfo {author} {\bibfnamefont {X.}~\bibnamefont
  {Pladevall}}, \bibinfo {author} {\bibfnamefont {X.}~\bibnamefont {Oriols}}, \
  and\ \bibinfo {author} {\bibfnamefont {J.}~\bibnamefont {Mompart}},\ }\href
  {http://books.google.it/books?id=mnqNx66amcIC} {\emph {\bibinfo {title}
  {Applied Bohmian Mechanics: From Nanoscale Systems to Cosmology}}}\ (\bibinfo
   {publisher} {Pan Stanford Publishing},\ \bibinfo {year} {2012})\BibitemShut
  {NoStop}%
\bibitem [{\citenamefont {Daumer}\ \emph {et~al.}(1997)\citenamefont {Daumer},
  \citenamefont {D{{\"u}}rr}, \citenamefont {Goldstein},\ and\ \citenamefont
  {Zanghi}}]{DaumerDurrGoldstein1997}%
  \BibitemOpen
  \bibfield  {author} {\bibinfo {author} {\bibfnamefont {M.}~\bibnamefont
  {Daumer}}, \bibinfo {author} {\bibfnamefont {D.}~\bibnamefont {D{{\"u}}rr}},
  \bibinfo {author} {\bibfnamefont {S.}~\bibnamefont {Goldstein}}, \ and\
  \bibinfo {author} {\bibfnamefont {N.}~\bibnamefont {Zanghi}},\ }\href
  {http://dx.doi.org/10.1023/B:JOSS.0000015181.86864.fb} {\bibfield  {journal}
  {\bibinfo  {journal} {Journal of Statistical Physics}\ }\textbf {\bibinfo
  {volume} {88}},\ \bibinfo {pages} {967} (\bibinfo {year} {1997})}\BibitemShut
  {NoStop}%
\bibitem [{\citenamefont {Ruggenthaler}\ \emph {et~al.}(2005)\citenamefont
  {Ruggenthaler}, \citenamefont {Gr{{\"u}}bl},\ and\ \citenamefont
  {Kreidl}}]{RuggenthalerGrublKreidl2005}%
  \BibitemOpen
  \bibfield  {author} {\bibinfo {author} {\bibfnamefont {M.}~\bibnamefont
  {Ruggenthaler}}, \bibinfo {author} {\bibfnamefont {G.}~\bibnamefont
  {Gr{{\"u}}bl}}, \ and\ \bibinfo {author} {\bibfnamefont {S.}~\bibnamefont
  {Kreidl}},\ }\href {http://stacks.iop.org/0305-4470/38/i=39/a=010} {\bibfield
   {journal} {\bibinfo  {journal} {Journal of Physics A: Mathematical and
  General}\ }\textbf {\bibinfo {volume} {38}},\ \bibinfo {pages} {8445}
  (\bibinfo {year} {2005})}\BibitemShut {NoStop}%
\bibitem [{\citenamefont {Wiseman}(2007)}]{Wiseman2007}%
  \BibitemOpen
  \bibfield  {author} {\bibinfo {author} {\bibfnamefont {H.~M.}\ \bibnamefont
  {Wiseman}},\ }\href {http://stacks.iop.org/1367-2630/9/i=6/a=165} {\bibfield
  {journal} {\bibinfo  {journal} {New Journal of Physics}\ }\textbf {\bibinfo
  {volume} {9}},\ \bibinfo {pages} {165} (\bibinfo {year} {2007})}\BibitemShut
  {NoStop}%
\bibitem [{\citenamefont {Kocsis}\ \emph {et~al.}(2011)\citenamefont {Kocsis},
  \citenamefont {Braverman}, \citenamefont {Ravets}, \citenamefont {Stevens},
  \citenamefont {Mirin}, \citenamefont {Shalm},\ and\ \citenamefont
  {Steinberg}}]{KocsisBravermanRavets2011}%
  \BibitemOpen
  \bibfield  {author} {\bibinfo {author} {\bibfnamefont {S.}~\bibnamefont
  {Kocsis}}, \bibinfo {author} {\bibfnamefont {B.}~\bibnamefont {Braverman}},
  \bibinfo {author} {\bibfnamefont {S.}~\bibnamefont {Ravets}}, \bibinfo
  {author} {\bibfnamefont {M.~J.}\ \bibnamefont {Stevens}}, \bibinfo {author}
  {\bibfnamefont {R.~P.}\ \bibnamefont {Mirin}}, \bibinfo {author}
  {\bibfnamefont {L.~K.}\ \bibnamefont {Shalm}}, \ and\ \bibinfo {author}
  {\bibfnamefont {A.~M.}\ \bibnamefont {Steinberg}},\ }\href {\doibase
  10.1126/science.1202218} {\bibfield  {journal} {\bibinfo  {journal}
  {Science}\ }\textbf {\bibinfo {volume} {332}},\ \bibinfo {pages} {1170}
  (\bibinfo {year} {2011})},\ \Eprint
  {http://arxiv.org/abs/http://www.sciencemag.org/content/332/6034/1170.full.pdf}
  {http://www.sciencemag.org/content/332/6034/1170.full.pdf} \BibitemShut
  {NoStop}%
\bibitem [{\citenamefont {Traversa}\ \emph {et~al.}(2013)\citenamefont
  {Traversa}, \citenamefont {Albareda}, \citenamefont {Di~Ventra},\ and\
  \citenamefont {Oriols}}]{TraversaAlbaredaDi-Ventra2012}%
  \BibitemOpen
  \bibfield  {author} {\bibinfo {author} {\bibfnamefont {F.~L.}\ \bibnamefont
  {Traversa}}, \bibinfo {author} {\bibfnamefont {G.}~\bibnamefont {Albareda}},
  \bibinfo {author} {\bibfnamefont {M.}~\bibnamefont {Di~Ventra}}, \ and\
  \bibinfo {author} {\bibfnamefont {X.}~\bibnamefont {Oriols}},\ }\href@noop {}
  {\bibfield  {journal} {\bibinfo  {journal} {Phys. Rev. A}\ }\textbf {\bibinfo
  {volume} {87}},\ \bibinfo {pages} {052124} (\bibinfo {year} {2013})},\
  \Eprint {http://arxiv.org/abs/1211.2357} {arXiv:1211.2357 [quant-ph]}
  \BibitemShut {NoStop}%
\bibitem [{\citenamefont {Brenig}\ and\ \citenamefont
  {Haag}(1959)}]{BrenigHaag1959}%
  \BibitemOpen
  \bibfield  {author} {\bibinfo {author} {\bibfnamefont {W.}~\bibnamefont
  {Brenig}}\ and\ \bibinfo {author} {\bibfnamefont {R.}~\bibnamefont {Haag}},\
  }\href {\doibase 10.1002/prop.19590070402} {\bibfield  {journal} {\bibinfo
  {journal} {Fortschritte der Physik}\ }\textbf {\bibinfo {volume} {7}},\
  \bibinfo {pages} {183} (\bibinfo {year} {1959})}\BibitemShut {NoStop}%
\bibitem [{\citenamefont {Dollard}(1969)}]{Dollard1969}%
  \BibitemOpen
  \bibfield  {author} {\bibinfo {author} {\bibfnamefont {J.}~\bibnamefont
  {Dollard}},\ }\href {\doibase 10.1007/BF01661573} {\bibfield  {journal}
  {\bibinfo  {journal} {Communications in Mathematical Physics}\ }\textbf
  {\bibinfo {volume} {12}},\ \bibinfo {pages} {193} (\bibinfo {year}
  {1969})}\BibitemShut {NoStop}%
\bibitem [{\citenamefont {D{{\"u}}rr}\ and\ \citenamefont
  {Teufel}(2009)}]{DurrTeufel2009}%
  \BibitemOpen
  \bibfield  {author} {\bibinfo {author} {\bibfnamefont {D.}~\bibnamefont
  {D{{\"u}}rr}}\ and\ \bibinfo {author} {\bibfnamefont {S.}~\bibnamefont
  {Teufel}},\ }\href {http://books.google.it/books?id=UWP2ZSs-UD0C} {\emph
  {\bibinfo {title} {Bohmian mechanics: the physics and mathematics of quantum
  theory}}},\ Fundamental Theories of Physics\ (\bibinfo  {publisher}
  {Springer},\ \bibinfo {year} {2009})\BibitemShut {NoStop}%
\bibitem [{\citenamefont {Yearsley}\ \emph {et~al.}(2011)\citenamefont
  {Yearsley}, \citenamefont {Downs}, \citenamefont {Halliwell},\ and\
  \citenamefont {Hashagen}}]{YearsleyDownsHalliwell2011}%
  \BibitemOpen
  \bibfield  {author} {\bibinfo {author} {\bibfnamefont {J.~M.}\ \bibnamefont
  {Yearsley}}, \bibinfo {author} {\bibfnamefont {D.~A.}\ \bibnamefont {Downs}},
  \bibinfo {author} {\bibfnamefont {J.~J.}\ \bibnamefont {Halliwell}}, \ and\
  \bibinfo {author} {\bibfnamefont {A.~K.}\ \bibnamefont {Hashagen}},\ }\href
  {\doibase 10.1103/PhysRevA.84.022109} {\bibfield  {journal} {\bibinfo
  {journal} {Phys. Rev. A}\ }\textbf {\bibinfo {volume} {84}},\ \bibinfo
  {pages} {022109} (\bibinfo {year} {2011})}\BibitemShut {NoStop}%
\end{thebibliography}%

\end{document}